\documentstyle[preprint,aps,version2]{revtex}
\begin{document}

\begin{title}
Positivity  restrictions to the transverse polarization 
of the inclusively detected  spin-half  baryons  in 
unpolarized electron-positron annihilation 
\end{title}

\author{Chun-Gui Duan$^{1,2,3}$ and Wei Lu$^{1,2}$ }

\begin{instit}
1. CCAST (World Laboratory), P.O. Box 8730, Beijing 100080, China 
 
2. Institute of High Energy Physics, 
 P.O. Box 918(4), Beijing 100039, China\footnote{Mailing address}

3. Department of Physics, Hebei Normal University, 
Hebei 050016, China 

 \end{instit}

\begin{abstract}

  The positivity  constraints  to the structure functions 
for the inclusive spin-half baryon  production by a time-like 
photon fragmentation are investigated. One conclusion is that $\hat F$,
 which arises from the  hadronic final-state interactions,
is  subjected to  an  inequality between its absolute 
value and the two spin-independent structure functions. 
On the basis of this finding, we derive a formula 
through which the upper limits can be given  for the 
transverse polarization of the inclusively detected spin-half 
baryons  in unpolarized electron-positron annihilation. 
The derived upper bound  supplies a consistency check 
for the judgement of reliability of experimental data and 
model calculations. 
\end {abstract}

\noindent{PACS Numbers: 13.88.+e, 13.65.+i}

\newpage

 Since the Fermilab's pioneering discovery \cite{Bunce} 
of a large transverse 
polarization of the inclusively detected $\Lambda$ hyperons, 
it  has been a challenge  for the 
particle physics community to understand such an 
observation.  Recently, one of the authors
 pointed out \cite{lw} that  there 
exists the possibility to approach alternatively transverse 
polarization phenomena  in unpolarized electron-positron 
annihilation.  The basic point is that, due to the 
final-state interactions, the decomposition of the 
time-like photon fragmentation 
tensor for the inclusive spin-half particle production 
contains a  component that is  odd under 
the naive time-reversal transformation.  The corresponding 
structure function $\hat F$ is directly related to the transverse 
polarization of the inclusively detected spin-half baryon
in unpolarized electron-positron annihilation. 
Since $\hat F$ is an experimentally accessible quantity, 
it is desirable to learn more about it before the relevant 
experiments come true.  Towards this direction,  
Lu, Li and Hu \cite{llh} worked out 
a  factorized expression of $\hat F$
in terms of the quark$\to$baryon 
fragmentation matrix elements. However, such an approach 
cannot provide us with any quantitative information about $\hat F$. 

  Transverse polarization of the inclusively detected baryons 
is essentially a one-power suppressed, namely, twist-three 
effects \cite{qiu}. Generally, one believes that it comes about  at 
lower energy regions and has no observable effects at high 
momentum transfers.  However, ALEPH Collaboration \cite{Aleph} 
measured  unexpectedly a few percent of  transverse 
component of the polarization of the inclusively detected 
$\Lambda$ hyperons at the $Z$ resonance.  This fact stimulates us 
to expect a sizable transverse polarization of the 
inclusively detected spin-half baryons at lower
c.m. energies of  unpolarized electron-positron annihilation. 
The corresponding experiments seem  to be beyond 
reach at present experimental facilities due to low 
statistics  but can be expected at future high-luminosity 
and moderate-energy electron-positron colliders such 
as  Beauty and Tau-Charm Factories.  Therefore, it 
is desirable to learn more systematics  about this 
transverse polarization phenomena before the relevant 
 experiments come true.

 In this paper, we  present a  positivity analysis 
of the hadronic tensor for the inclusive spin-half baryon 
production by a time-like photon, with a result that 
the upper limit is 
given for the considered transverse 
polarization.   Since the transverse polarization of 
the inclusive baryons is a twist-three effect, 
we work at the  moderate c.m. energies,  far 
bellow the $Z$ resonance.  Thus, 
the one-photon annihilation approximation is  well applicable, 
and the corresponding   hadronic tensor  is also termed 
the photon fragmentation tensor.

Under the one-photon approximation, the inclusive  
single-particle annihilation can be $naively$ taken 
as the crossed process of the  deeply inelastic scattering. 
Concerning the  latter, 
Doncel and de Rafael  \cite{nuovo} derived   early in 1970's  a series 
of  useful inequalities among its structure functions. 
However, the kinematics of the former is 
quite different from that of the latter. Furthermore, 
there exist strong  final-state interactions 
in the inclusively detected  hadronic final state, 
which cause one more structure function \cite{lw}. 
Hence, it  will be  necessary to  perform a positivity analysis 
on the   hadronic tensor for the inclusive spin-half baryon 
annihilation, with the polarization of the inclusive spin-half
baryon monitored.  Obviously, our results will be 
applicable to the inclusive single-pseudoscalar-meson annihilation.

What we  will consider is 
\begin{equation} 
e^- + e^+ \to \gamma^\ast(q) \to B(k, s) +X, 
\end{equation}
where $q$ and $k$ are the momenta of the 
corresponding particles and $s$ is the spin  four-vector of 
the inclusively detected spin-half baryon.
For a pure state of the inclusive spin-half baryon, 
we take the normalization 
\begin{equation} 
\langle k,s|k^\prime,s^\prime\rangle =(2\pi)^3 
2 E \delta^{3} ({\bf k} -{\bf k}^\prime) \delta_{s s^\prime}
{~~\rm with}~ s\cdot s=-1. 
\end{equation} 
As usual,  the  photon fragmentation tensor is defined as 
\begin{equation} 
\label{tensor}
\tilde W_{\mu\nu}(k,q,s)=\displaystyle\frac{1}{4\pi}
\sum\limits_X \int d^4 x \exp (i q\cdot x)
\langle 0|j_\mu(0)|B(k,s),X\rangle \langle B(k,s),X|j_\nu(x)|0\rangle ,
\end{equation} 
where $j^\mu= \sum_f e_f \bar \psi_f \gamma^\mu \psi_f $ is the
electromagnetic current, with $f$ being the quark flavor index 
and $e_f$ being the quark charge in unit of the electron charge. 
From the general symmetry analysis,  $\tilde W_{\mu\nu}(q,k,s)$ can be 
expressed in terms of  experimentally measurable structure functions. 
Due to the final-state interactions in the 
inclusively detected  $out$ state, there exists one more structure 
function in the general Lorentz expansion 
of  $\tilde W_{\mu\nu}(q,k,s)$ in comparison to its deep inelastic 
scattering counterpart. Subjected to all symmetry constraints, 
$\tilde W_{\mu\nu}(q,k,s)$  takes 
the following Lorentz decomposition \cite{lw} 
\begin{eqnarray}
\label{dec}
\tilde W_{\mu\nu}(k,q,s)&
=&\frac{1}{2}
\left[ (-g_{\mu\nu}+ \frac{q_\mu q_\nu}{Q^2})\tilde F_1(\nu,Q)
+(k_\mu - \frac{\nu}{Q^2}q_\mu) 
(k_\nu - 
\frac{\nu}{Q^2}q_\nu) \frac{\tilde F_2(\nu,Q)}{\nu} \right]
\nonumber  \\
& & +iM  \varepsilon_{\mu\nu\alpha\beta}q^\alpha s^\beta 
 \frac{\tilde g_1(\nu,Q)}{\nu}
+iM  \varepsilon_{\mu\nu\alpha\beta}q^\alpha 
(s^\beta -\frac{s\cdot q}{\nu} k^\beta) \frac{\tilde g_2(\nu,Q)}
{\nu}
\nonumber \\ 
& & 
+M  \left[(k_\mu- \frac{\nu}{Q^2}
q_\mu)\varepsilon_{\nu \alpha \beta \gamma} 
k^\alpha q^\beta s^\gamma 
+(k_\nu- \frac{\nu}{Q^2}q_\nu)
\varepsilon_{\mu \alpha \beta \gamma} 
k^\alpha q^\beta s^\gamma \right]  \frac{\hat F(\nu,Q)}{\nu^2},
\end{eqnarray}
where $k\cdot q=\nu$,  $Q=\sqrt{q ^2}$, 
and $M$ is the mass of the inclusive baryon. 
The structure function $\hat F$, which has no counterpart in the 
deeply inelastic scattering case,  arises from the 
final-state interaction, so it is not  forbidden 
by the time reversal invariance.

Our basic starting point  is  the Hermiticity of the electromagnetic 
current. For an arbitrary  Lorentz vector $a^\mu$, one can show  
\begin{equation} 
\tilde W_{\mu\nu}(q,k,s) a^{\ast \mu} a^\nu = \frac{1}{4\pi}
\sum\limits_X (2\pi)^4 \delta^4 (q-k-k_X)
|\langle B(k,s),X|a\cdot j|0\rangle |^2, 
\end{equation} 
so $\tilde W_{\mu\nu}(q,k,s)$ is semi-positive definite: 
\begin{equation} 
\tilde W_{\mu\nu}(q,k,s) a^{\ast \mu} a^\nu \geq 0. \label{poi}
\end{equation}

 Before addressing  the general constraints of the 
positivity on  the structure functions,  we 
first discuss  several   special cases  by specifying 
$a$ to be the photon polarization vectors $e_i$ and 
letting the inclusive 
spin-half baryon be in its  specific spin state. 
For this purpose,  we work in the c.m. frame, choosing the  
$\hat x$-$\hat z$ plane in the production plane with the $\hat z$ axis 
along the outgoing direction of the inclusive particle.  
Then, we introduce three polarization vectors for the virtual photon: 
\begin{equation} 
 e^\mu_1= -\frac{1}{\sqrt 2} (0, 1, +i, 0),
~ e^\mu_2= +\frac{1}{\sqrt 2} (0, 1, -i, 0),
~ e^\mu_3= (0, 0, 0, 1 ). 
\end{equation} 
Notice that all three polarization vectors are 
orthnormal, namely, 
\begin{equation} 
e^\ast_i \cdot e_j =-\delta_{ij}. 
\end{equation} 
In  addition, they satisfy the Lorentz condition 
\begin{equation} 
e_i\cdot q=0,  ~~i=1, 2, 3. 
\end{equation} 

Now we consider the case in which 
the inclusive  spin-half baryon  is polarized along its momentum 
direction. From (\ref{poi}), it is easy  to derive 
\begin{equation} \label{x1}
\frac{1}{2} \tilde F_1 -\tilde g_1 -\frac{M^2Q^2 }{\nu^2}\tilde g_2 \geq 0, 
\end{equation}
\begin{equation} \label{x2}
\frac{1}{2} \tilde F_1 +\tilde g_1 +\frac{M^2Q^2}{\nu^2}\tilde g_2 \geq 0, 
\end{equation}
and 
\begin{equation} \label{x3}
\tilde F_1 +\frac{\xi^2}{\nu Q^2} \tilde F_2 \geq 0, 
 ~~{\rm with~~} \xi=\sqrt{\nu^2-M^2Q^2}.
\end{equation}
These inequalities correspond to  $a=e_i$,  $i=1,2,3$, respectively.

Adding (\ref{x1}) to  (\ref{x2}), it results in 
\begin{equation} \label{15}
  \tilde F_1 \geq 0. 
\end{equation} 
On the other hand, (\ref{x1}) to  (\ref{x2}) can be 
synthesized into a more compact form: 
\begin{equation}\label{?} 
\frac{1}{2} \tilde F_1 \geq |\tilde g_1+\frac{M^2Q^2}{\nu^2} \tilde g_2|.
\end{equation} 
Here we note in passing that (\ref{x3}), (\ref{15}), and 
(\ref{?}) have their counterparts \cite{nuovo} in the deeply inelastic 
scattering, up to the kinematical  differences. 

By  choosing the inclusive spin-half baryon to 
be longitudinally polarized,  we  have not found 
any inequality concerning  $\hat F$ from the above discussion. 
To understand  such  a fact, we need only 
to  remind that $\hat F$ is related to the transverse polarization 
of the inclusive spin-half baryon.  So, we are encouraged to 
consider the case in which 
the spin of the inclusive spin-half baryon is 
aligned along  the $\hat y$ axis, i. e., parallel to 
the normal of the production plane.  However,  the 
result is again discouraging. What is worse, in this case  we cannot 
obtain  (\ref{x1}) and (\ref{x2}), whereas (\ref{x3}) and (\ref{15})
are reproduced. 

At this stage, we  can conclude that 
the positivity constraints obtained by considering  some 
specific spin states of the inclusive spin-half particles, 
albeit correct, are incomplete. 
The underlying reason is that  all the above  inequalities 
 being the  direct consequences of  (\ref{poi}),  are  only 
the necessary conditions for the positivity of the hadronic tensor.  
However, they  are not the sufficient conditions. 

In the following, we will present a complete
analysis  of the sufficient conditions 
on the basis of the spin density matrix description. 
In general, the cross section for the  process
 considered can be expressed as 
\begin{equation} 
\frac{d \sigma}{d E d \cos\theta} = 
\frac{4 \pi \alpha^2}{Q^5}L_{\mu\nu} \tilde W^{\mu\nu}_{\lambda\lambda^\prime}
\rho_{\lambda^\prime \lambda},
\end{equation} 
where $\rho_{\lambda^\prime \lambda}$ is the spin  density matrix 
of the inclusive spin-half baryon, and 
$\tilde W^{\mu\nu}_{\lambda\lambda^\prime} (q,k)$ is the hadronic tensor 
defined as 
\begin{equation} 
\label{lambda}
\tilde W^{\mu\nu}_{\lambda\lambda^\prime}   
(k,q)=\displaystyle\frac{1}{4\pi}
\sum\limits_X \int d^4 x \exp (i q\cdot x)
\langle 0|j^\mu(0)|B(k,\lambda),X\rangle \langle B(k,\lambda^\prime),X|j^\nu(x)|0\rangle .
\end{equation} 
In terms of $\tilde W^{\mu\nu}_{\lambda\lambda^\prime} (q,k)$,  the 
positivity condition  can be written as 
\begin{equation}
\sum\limits_{\lambda\lambda^\prime} 
\tilde W^{\mu\nu}_{\lambda\lambda^\prime} (q,k) A_{\mu\lambda}A^\ast_{\nu
\lambda^\prime} \geq 0, 
\end{equation} 
where $A^\mu_\lambda$  can be any complex vector 
in the eight-dimensional space  spanned by  the Lorentz indices 
$\mu=0,1,2,3$ and  spin  indices $\lambda=+\frac{1}{2}, 
-\frac{1}{2}$.  

 Considering that the spin vector satisfies $s\cdot k$=0,  we 
introduce   the following three  orthonormal four-vectors: 
\begin{equation} 
u^\mu=(0,-1,0, 0), ~v^\mu=(0,0,+1, 0),
~ w^\mu=\frac{1}{MQ} (\xi, 0,0, \nu). 
\end{equation} 
Obviously, these four-vectors are orthogonal to $k$,  the momentum 
of the inclusive particle.  Using $u$, $v$ and $w$, we can 
most generally  express $\tilde W^{\mu\nu}_{\lambda\lambda^\prime} (q,k)$ as 

\begin{eqnarray}
\label{deco}
\tilde W_{\mu\nu, \lambda\lambda^\prime}(k,q)&
=&\frac{1}{2}\delta_{\lambda\lambda^\prime} 
\left[ (-g_{\mu\nu}+ \frac{q_\mu q_\nu}{Q^2})\tilde F_1(\nu,Q)
+(k_\mu - \frac{\nu}{Q^2}q_\mu) 
(k_\nu - 
\frac{\nu}{Q^2}q_\nu) \frac{\tilde F_2(\nu,Q)}{\nu} \right]
\nonumber  \\
& & + 
\left( u^\rho (\tau_1)_{\lambda\lambda^\prime} 
+v^\rho (\tau_2)_{\lambda\lambda^\prime} 
+w^\rho (\tau_3)_{\lambda\lambda^\prime} \right)
\left\{ \frac{iM }{\nu} \varepsilon_{\mu\nu\alpha\rho}q^\alpha 
(\tilde g_1(\nu,Q) + \tilde g_2(\nu,Q) ) 
 \right. 
\nonumber \\ 
& & 
+\left. \frac{M}{\nu^2}  \left[(k_\mu- \frac{\nu}{Q^2}
q_\mu)\varepsilon_{\nu \alpha \beta\rho} 
k^\alpha q^\beta  
+(k_\nu- \frac{\nu}{Q^2}q_\nu)
\varepsilon_{\mu \alpha \beta \rho } 
k^\alpha q^\beta \right] \hat F(\nu,Q) \right\}
\nonumber \\ 
& & - \frac{ iM}{\nu^2}   \varepsilon_{\mu\nu\alpha\beta}q^\alpha k^\beta 
q_\rho\left( u^\rho (\tau_1)_{\lambda\lambda^\prime} 
+v^\rho (\tau_2)_{\lambda\lambda^\prime} 
+w^\rho (\tau_3)_{\lambda\lambda^\prime} \right)
\tilde g_2(\nu,Q), 
\end{eqnarray}
where $\tau_{1,2,3}$ are the Pauli matrices. 

    Notice that $\tilde W_{\mu\nu,\lambda\lambda^\prime}(q,k)$ 
is an eight-dimensional matrix and we can simply  put it into 
the form 

\begin{equation} 
\tilde W_{\mu\nu}(\nu,Q^2)_{\lambda,\lambda^\prime}= 
\left(
\begin{array}{cc} 
\tilde W_{\mu\nu}(\nu,Q^2)_{+\frac{1}{2},+\frac{1}{2}}&
\tilde W_{\mu\nu}(\nu,Q^2)_{+\frac{1}{2},-\frac{1}{2}}\\
\tilde W_{\mu\nu}(\nu,Q^2)_{-\frac{1}{2},+\frac{1}{2}}&
\tilde W_{\mu\nu}(\nu,Q^2)_{-\frac{1}{2},-\frac{1}{2}}
\end{array}
\right).  \label{direct1} 
\end{equation} 
In our kinematics, the four submatrices reads 
\begin{equation} 
\tilde W_{\mu\nu}(\nu,Q^2)_{\pm\frac{1}{2},\pm\frac{1}{2}}=
\left(
\begin{array}{cccc} 
0 & 0 & 0 & 0 \\
0 & \frac{1}{2}\tilde F_1  & \pm i A & 0 \\
0 & \mp iA & \frac{1}{2}\tilde F_1  & 0 \\
0 & 0 & 0 & \frac{1}{2}\tilde F_1 +\frac{\xi^2}{2\nu Q^2}\tilde F_2
\end{array}
\right),  \label{cat1} 
\end{equation} 
and 
\begin{equation} 
\tilde W_{\mu\nu}(\nu,Q^2)_{\pm \frac{1}{2},\mp \frac{1}{2}}=
\left(
\begin{array}{cccc} 
0 & 0 & 0 & 0 \\
0 & 0 & 0 & 
\mp iB^\ast 
\\
0 & 0 & 0 & B^\ast \\
0 & \mp i B  &   B  & 0
\end{array}
\right), \label{cat2}
\end{equation}
where 
\begin{equation}
A=\tilde g_1 +\frac{M^2Q^2}{\nu^2}\tilde g_2, 
\end{equation}
\begin{equation}
B=\frac{iMQ}{\nu}(\tilde g_1+\tilde g_2)+\frac{M\xi^2}{Q\nu^2}\hat F. 
\end{equation}
Obviously,   $\tilde W_{\mu\nu}(\nu,Q^2)_{\lambda,\lambda^\prime}$ satisfies 
the following  Hermiticity relation: 
\begin{equation}
[\tilde W_{\mu\nu}(\nu,Q^2)_{\lambda,\lambda^\prime}]^\ast 
=
\tilde W_{\nu\mu}(\nu,Q^2)_{\lambda^\prime,\lambda}. 
\end{equation}
Owing to the current conservation condition, 
each submatrix at fixed $\lambda$ and $\lambda^\prime$ 
is  at most  of rank three, which can be easily seen 
in Eqs. (\ref{cat1}) and (\ref{cat2}) by eliminating the 
first column and first row entries.  After doing so, 
we are left with four $3\times 3$ submatrices.

Notice that the necessary and  sufficient conditions  
for  a matrix  to be  semi-positive finite are that 
all its submatrices  have a  semi-positive determinant \cite{soffer}. 
Exhausting all the possibilities for 
$\tilde W_{\mu\nu}(k,q)_{\lambda\lambda^\prime}$, 
one can gain directly the restrictions 
stated by (\ref{x1}), (\ref{x2}) and (\ref{x3}). 
Furthermore, there exists the following inequality 
\begin{equation} 
\frac{M^2Q^2}{\nu^2}(\tilde g_1 +\tilde g_2)^2
+\frac{M^2\xi^4}{Q^2\nu^4} \hat  F^2 \leq \frac{1}{4}
\tilde F_1 (\tilde F_1 + \frac{\xi^2}{\nu Q^2} \tilde F_2). 
\label{new}
\end{equation} 
Nevertheless, there is no  further restrictions.  From 
(\ref{new}),  one can deduce the following  two  inequalities: 
\begin{equation} 
|\tilde g_1 +\tilde g_2|
\leq \frac{\nu}{2 M Q}
\sqrt{ \tilde F_1 (\tilde F_1 + \frac{\xi^2}{\nu Q^2} \tilde F_2)},
\label{yy1}
\end{equation} 

\begin{equation} 
|\hat  F| 
\leq \frac{Q\nu^2}{2 M \xi^2}
\sqrt{ \tilde F_1 (\tilde F_1 + \frac{\xi^2}{\nu Q^2} \tilde F_2)}. 
\label{yy2}
\end{equation} 
Therefore, the complete  positivity constraints on the 
inclusive spin-half  baryon  annihilation  include
(\ref{x3}), (\ref{15}), (\ref{?}), (\ref{yy1}) and (\ref{yy2}). 

Since $\hat F$  is related directly to the transverse  polarization 
of the inclusive spin-half baryon, from (\ref{yy2})
we  can obtain some upper limits  for the absolute of the 
transverse polarization. In doing so, 
we need to make use of the following Callan-Gross relation: 
\begin{equation} 
 \tilde F_1 +  \frac{E}{Q} \tilde F_2 =0,
\label{callan} 
\end{equation} 
which  can be justified  \cite{lw2} both in the quark-parton model 
and at the  leading-twist factorization.  The modifications to 
Eq. (\ref{callan})  come about at twist four. 
Since  the transverse polarization is at twist three,  we 
can safely  employ Eq. (\ref{callan}) in the present work.

Experimentally,  one measures the energy 
$E$ of the inclusive particle and its outgoing angle  $\theta$ 
with respect to the beam direction. In terms of these variables, 
the transverse polarization can be put into \cite{lw}  
\begin{equation} 
P(Q,E,\theta)= 
 \frac{4M  (E^2 -M^2) \sin\theta \cos\theta\hat F}
{E \left[2 EQ \tilde F_1 +(E^2-M^2) \sin^2\theta \tilde F_2\right]}.
\label{lw}
\end{equation} 
Substituting (\ref{yy2}) into 
(\ref{lw}) and utilizing (\ref{callan}), one has 
\begin{equation} 
|P(Q, E, \theta)|\leq \left|
 \frac{4M E \sin\theta \cos\theta} 
{E^2(1+\cos^2\theta)+M^2\sin^2\theta}\right|,
\label{ine}
\end{equation} 
from which  the upper limits can be given of the transverse polarization. 
It should be noted  that one cannot determine the sign of 
transverse polarization from the positivity condition of the 
photon fragmentation tensor.

 The implications of the derived positivity restrictions 
are two-fold.  On the one hand, they can  exclude some 
kinematical regions in which to observe large transverse 
polarization. On the other hand,  they  supply 
us an  important benchmark to judge the reasonability and 
reliability of the  experimental data on transverse 
polarization, especially when 
there exist large statistical errors. 
 To  acquire a quantitative sense, let  us consider  the 
inclusive $\Lambda$ hyperon production at the future 
Tau-Charm factories \cite{tau}.  Supposing such a machine is operated 
at  a c.m. energy of $\sqrt{s}=4.25$ GeV,   the maximum energy of 
the $\Lambda$ particle will be $E_{\rm max}= 
\frac{1}{2}\sqrt{s-4M^2_\Lambda}=1.81$ GeV, which 
corresponds to the $\bar\Lambda\Lambda$ pair production. 
After a little algebra,  one can  know that the most optimal 
upper bound for $|P|$  is 
$37\%$ as $\theta\approx 50^\circ$. 

 In  summary, we examined the  positivity constraints 
to the structure functions for the inclusive spin-half baryon 
production by a time-like photon fragmentation. 
We found that 
the  absolute value $|\hat F|$ of the 
final-state-interaction-caused structure function is 
smaller than  a combination of $\tilde F_1$ and $\tilde F_2$,
two spin-independent structure functions. 
As s result, we deduce an upper limit for the 
transverse polarization of the inclusively detected spin-
half baryons in unpolarized electron-positron annihilation, 
which can   serve as a consistency check of  the reliability 
of experimentally data and model calculations in the future.

{\it Acknowledgement} One of the authors (W. L.) 
 would like to thank 
Professor J. Soffer for useful correspondence about the 
positivity.

\end{document}